\documentclass[twocolumn,showpacs,aps,prl,superscriptaddress]{revtex4}

\usepackage{graphicx}
\usepackage{dcolumn}
\usepackage{bm}

\begin{document}

\title{Strange Quark Contributions to Parity-Violating Asymmetries in the Backward Angle G0 Electron Scattering Experiment
}

\author{D.~Androi\'c}
\affiliation{Department of Physics, University of Zagreb, Zagreb HR-41001 Croatia} 

\author{D.~S.~Armstrong}
\affiliation{Department of Physics, College of William and Mary, Williamsburg, VA 23187 USA} 

\author{J.~Arvieux$^\dagger$}
\affiliation{Institut de Physique Nucl\'eaire d'Orsay, Universit\'e Paris-Sud, F-91406 Orsay Cedex FRANCE}



\author{S.~L.~Bailey}
\affiliation{Department of Physics, College of William and Mary, Williamsburg, VA 23187 USA} 


\author{D.~H.~Beck}
\affiliation{Loomis Laboratory of Physics, University of Illinois, Urbana, IL 61801 USA}

\author{E.~J.~Beise}
\affiliation{Physics Department, University of Maryland, College Park, MD 20742 USA}

\author{J.~Benesch}
\affiliation{Thomas Jefferson National Accelerator Facility, Newport News, VA 23606 USA}

\author{F.~Benmokhtar}
\affiliation{Physics Department, University of Maryland, College Park, MD 20742 USA}
\affiliation {Department of Physics, Carnegie Mellon University, Pittsburgh, PA 15213 USA}

\author{L.~Bimbot}
\affiliation{Institut de Physique Nucl\'eaire d'Orsay, Universit\'e Paris-Sud, F-91406 Orsay Cedex FRANCE}

\author{J.~Birchall}
\affiliation{Department of Physics, University of Manitoba, Winnipeg, MB R3T 2N2 CANADA}


\author{P.~Bosted}
\affiliation{Thomas Jefferson National Accelerator Facility, Newport News, VA 23606 USA}


\author{H.~Breuer}
\affiliation{Physics Department, University of Maryland, College Park, MD 20742 USA}



\author{C.~L.~Capuano}
\affiliation{Department of Physics, College of William and Mary, Williamsburg, VA 23187 USA} 


\author{Y.-C.~Chao}
\affiliation{Thomas Jefferson National Accelerator Facility, Newport News, VA 23606 USA}




\author{A.~Coppens}
\affiliation{Department of Physics, University of Manitoba, Winnipeg, MB R3T 2N2 CANADA} 
 


\author{C.~A.~Davis}
\affiliation{TRIUMF, Vancouver, BC V6T 2A3 CANADA}

\author{C.~Ellis}
\affiliation{Physics Department, University of Maryland, College Park, MD 20742 USA}



\author{G.~Flores}
\affiliation{Physics Department, New Mexico State University, Las Cruces, NM 88003 USA}


\author{G.~Franklin}
\affiliation {Department of Physics, Carnegie Mellon University, Pittsburgh, PA 15213 USA}

\author{C.~Furget}
\affiliation{LPSC, Universit\'e Joseph Fourier Grenoble 1, CNRS/IN2P3, Institut  Polytechnique de Grenoble, Grenoble, FRANCE}

\author{D.~Gaskell}
\affiliation{Thomas Jefferson National Accelerator Facility, Newport News, VA 23606 USA}

\author{M.~T.~W.~Gericke}
\affiliation{Department of Physics, University of Manitoba, Winnipeg, MB R3T 2N2 CANADA}

\author{J.~Grames}
\affiliation{Thomas Jefferson National Accelerator Facility, Newport News, VA 23606 USA}



\author{G.~Guillard}
\affiliation{LPSC, Universit\'e Joseph Fourier Grenoble 1, CNRS/IN2P3, Institut  Polytechnique de Grenoble, Grenoble, FRANCE}





\author{J.~Hansknecht}
\affiliation{Thomas Jefferson National Accelerator Facility, Newport News, VA 23606 USA}



\author{T.~Horn}
\affiliation{Thomas Jefferson National Accelerator Facility, Newport News, VA 23606 USA}


\author{M.~Jones}
\affiliation{Thomas Jefferson National Accelerator Facility, Newport News, VA 23606 USA}



\author{P.~M.~King}
\affiliation{Department of Physics and Astronomy, Ohio University, Athens, OH 45701 USA}



\author{W.~Korsch}
\affiliation{Department of Physics and Astronomy, University of Kentucky, Lexington, KY 40506 USA}

\author{S.~Kox}
\affiliation{LPSC, Universit\'e Joseph Fourier Grenoble 1, CNRS/IN2P3, Institut  Polytechnique de Grenoble, Grenoble, FRANCE}



\author{L.~Lee}
\affiliation{Department of Physics, University of Manitoba, Winnipeg, MB R3T 2N2 CANADA}



\author{J.~Liu}
\affiliation{Kellogg Radiation Laboratory, California Institute of Technology,  Pasadena, CA 91125 USA}


\author{A.~Lung}
\affiliation{Thomas Jefferson National Accelerator Facility, Newport News, VA 23606 USA}


\author{J.~Mammei}
\affiliation{Department of Physics, Virginia Tech, Blacksburg, VA 24061 USA}


\author{J.~W.~Martin}
\affiliation{Department of Physics, University of Winnipeg, Winnipeg, MB R3B 2E9 CANADA}



\author{R.~D.~McKeown}
\affiliation{Kellogg Radiation Laboratory, California Institute of Technology,  Pasadena, CA 91125 USA}


\author{M.~Mihovilovic}
\affiliation{Jo\^zef Stefan Institute, 1000 Ljubljana, SLOVENIA}

\author{A.~Micherdzinska}
\affiliation{Department of Physics, University of Winnipeg, Winnipeg, MB R3B 2E9 CANADA}

\author{H.~Mkrtchyan}
\affiliation{Yerevan Physics Institute, Yerevan 375036 ARMENIA}



\author{M.~Muether}
\affiliation{Loomis Laboratory of Physics, University of Illinois, Urbana, IL 61801 USA}







\author{S.~A.~Page}
\affiliation{Department of Physics, University of Manitoba, Winnipeg, MB R3T 2N2 CANADA}

\author{V.~Papavassiliou}
\affiliation{Physics Department, New Mexico State University, Las Cruces, NM 88003 USA}

\author{S.~F.~Pate}
\affiliation{Physics Department, New Mexico State University, Las Cruces, NM 88003 USA}

\author{S.~K.~Phillips}
\affiliation{Department of Physics, College of William and Mary, Williamsburg, VA 23187 USA} 

\author{P. Pillot}
\affiliation{LPSC, Universit\'e Joseph Fourier Grenoble 1, CNRS/IN2P3, Institut  Polytechnique de Grenoble, Grenoble, FRANCE}

\author{M.~L.~Pitt}
\affiliation{Department of Physics, Virginia Tech, Blacksburg, VA 24061 USA}

\author{M.~Poelker}
\affiliation{Thomas Jefferson National Accelerator Facility, Newport News, VA 23606 USA}



\author{B.~Quinn}
\affiliation {Department of Physics, Carnegie Mellon University, Pittsburgh, PA 15213 USA}

\author{W.~D.~Ramsay}
\affiliation{Department of Physics, University of Manitoba, Winnipeg, MB R3T 2N2 CANADA}


\author{J.-S.~Real}
\affiliation{LPSC, Universit\'e Joseph Fourier Grenoble 1, CNRS/IN2P3, Institut  Polytechnique de Grenoble, Grenoble, FRANCE}

\author{J.~Roche}
\affiliation{Department of Physics and Astronomy, Ohio University, Athens, OH 45701 USA}

\author{P.~Roos}
\affiliation{Physics Department, University of Maryland, College Park, MD 20742 USA}



\author{J.~Schaub}
\affiliation{Physics Department, New Mexico State University, Las Cruces, NM 88003 USA}

\author{T.~Seva}
\affiliation{Department of Physics, University of Zagreb, Zagreb HR-41001 Croatia} 

\author{N.~Simicevic}
\affiliation{Department of Physics, Louisiana Tech University,  Ruston, LA 71272 USA}

\author{G.~R.~Smith}
\affiliation{Thomas Jefferson National Accelerator Facility, Newport News, VA 23606 USA}

\author{D.~T.~Spayde}
\affiliation{Department of Physics, Hendrix College, Conway, AR 72032 USA}


\author{M.~Stutzman}
\affiliation{Thomas Jefferson National Accelerator Facility, Newport News, VA 23606 USA}

\author{R.~Suleiman}
\affiliation{Department of Physics, Virginia Tech, Blacksburg, VA 24061 USA}
\affiliation{Thomas Jefferson National Accelerator Facility, Newport News, VA 23606 USA}


\author{V.~Tadevosyan}
\affiliation{Yerevan Physics Institute, Yerevan 375036 ARMENIA}



\author{W.~T.~H.~van~Oers}
\affiliation{Department of Physics, University of Manitoba, Winnipeg, MB R3T 2N2 CANADA}

\author{M.~Versteegen}
\affiliation{LPSC, Universit\'e Joseph Fourier Grenoble 1, CNRS/IN2P3, Institut  Polytechnique de Grenoble, Grenoble, FRANCE}

\author{E.~Voutier}
\affiliation{LPSC, Universit\'e Joseph Fourier Grenoble 1, CNRS/IN2P3, Institut  Polytechnique de Grenoble, Grenoble, FRANCE}

\author{W.~Vulcan}
\affiliation{Thomas Jefferson National Accelerator Facility, Newport News, VA 23606 USA}


\author{S.~P.~Wells}
\affiliation{Department of Physics, Louisiana Tech University,  Ruston, LA 71272 USA}

\author{S.~E.~Williamson}
\affiliation{Loomis Laboratory of Physics, University of Illinois, Urbana, IL 61801 USA}

\author{S.~A.~Wood}
\affiliation{Thomas Jefferson National Accelerator Facility, Newport News, VA 23606 USA}




\collaboration{G0 Collaboration}


\noaffiliation

\date{\today}

\begin{abstract}
We have measured parity-violating asymmetries in elastic electron-proton and quasi-elastic electron-deuteron scattering at $Q^2 = 0.22$ and $0.63$ GeV$^2$.  They are sensitive to strange quark contributions to currents in the nucleon, and to the nucleon axial current.  The results indicate strange quark contributions of $\lesssim 10$\% of the charge and magnetic nucleon form factors at these four-momentum transfers.  
We also present the first measurement of anapole moment effects in the axial current at these four-momentum transfers.
\end{abstract}


\maketitle
 
At short distance scales, bound systems of quarks have relatively simple properties and QCD is successfully described by perturbation theory.  However, on the size scale of the bound state, $\sim 1$ fm, the QCD coupling constant is large and the effects of the color fields are a significant challenge, even in lattice QCD.  In addition to valence quarks, {\it e.g.}, $uud$ for the proton, there is a sea of gluons and $q \bar q$ pairs that plays an important role.  From a series of experiments measuring the parity-violating asymmetries of electrons scattered from protons and neutrons, 
we can extract 
the contributions of strange quarks to nucleon ground state charge and magnetic form factors.  These strange quark contributions are exclusively part of the quark sea because there are no strange valence quarks in the nucleon.


The SAMPLE~\cite{beisepittspayde}, HAPPEx~\cite{happex}, PVA4~\cite{mainz09} and G0~\cite{forward} experiments have previously reported measurements of these parity-violating asymmetries.
Using the combined forward angle asymmetries and the SAMPLE backward angle proton and deuteron measurements, a complete experimental determination of the strange quark vector currents and the axial current (see discussion below) has been made at a four-momentum transfer $Q^2 = 0.1$ GeV$^2$~\cite{liu}.
In this paper, we report the first complete backward angle asymmetry measurements since the SAMPLE experiment, at the four-momentum transfers of 0.221 and 0.628 GeV$^2$.  Together with our forward angle measurements~\cite{forward}, they allow the first experimental separation of these effects for $Q^2>0.1$ GeV$^2$.

For longitudinally polarized electrons ($R$ and $L$) scattered elastically from unpolarized protons (neutrons), the asymmetry is~\cite{beckmckeown}
\begin{widetext}
\begin{equation}
A = {d\sigma_R - d\sigma_L \over d\sigma_R + d\sigma_L}
= - \frac{G_F Q^2}{4 \sqrt{2} \pi \alpha}
{{
\varepsilon G^{\gamma}_{{E}} G^{Z}_
{{E}} + \tau G^{\gamma}_{{M}}
G^{Z}_{{M}} - (1-4 \sin^2 \theta_W ) 
\varepsilon^{\prime} G^{\gamma}_{{M}} G^{e}_{A}}   \over
{\varepsilon (G^\gamma_{E})^2 + \tau (G^\gamma_{M})^2}}
\label{eqn:asym}
\end{equation}
\end{widetext}
where $\tau = {{Q^2} / {4 M^2}}$, $\varepsilon = 1/\left (1 + 2(1 + \tau)\tan^2{\theta / 2} \right )$, $\varepsilon^{\prime} = \sqrt{\tau (1+\tau) (1- \varepsilon^2)}$, 
$Q^2$ is the squared four-momentum transfer ($Q^2 > 0$), $G_F$ and $\alpha$ the usual weak and electromagnetic couplings, $\theta_W$ the weak mixing angle, $\theta$ the laboratory electron scattering angle, $M$ the proton (neutron) mass and $G^{\gamma}_{{E}}$, $G^{Z}_{{E}}$, etc.\ the proton (neutron) electromagnetic and neutral weak form factors, respectively.  We measure quasi-elastic scattering from the deuteron where the asymmetry is, to a good approximation, the sum of those for the proton and neutron.  For the results reported here, we use a complete model of the electroweak deuteron response~\cite{schiavilla}. 

Separation of the strange quark contributions to nucleon currents 
was developed by Kaplan and Manohar~\cite{kaplan88a}.  Because the coupling of both photons and $Z$ bosons to point-like quarks is well defined, it is possible to separate the contributions of the various flavors by comparing the corresponding currents.  Neglecting the very small contribution from heavier flavors, the charge and magnetic form factors of the proton and neutron can be written ($i=\gamma,Z$) 
\begin{eqnarray}
\label{eqn:protonff}
G_{E,M}^{p,i}= e^{i,u} G_{E,M}^u + e^{i,d} \left( G_{E,M}^d + G_{E,M}^s \right), \\
G_{E,M}^{n,i}= e^{i,u} G_{E,M}^d + e^{i,d} \left( G_{E,M}^u + G_{E,M}^s \right), \nonumber
\end{eqnarray}
assuming the proton and neutron are related by a simple exchange of $u$ and $d$ quarks (as well as $\bar u$ and $\bar d$)~\cite{isospin}.  For the ordinary electromagnetic form factors the charges are $e^\gamma = +2/3$, $-1/3$ for $u$ and $d/s$ quarks, respectively.
Separating the contributions of the flavors and, in particular, isolating $G_{E,M}^s$, requires a third pair of observables -- the $G_{E,M}^{p,Z}$ that appear in the proton asymmetry above.  These form factors are written (Eq.\ (\ref{eqn:protonff})) in terms of the weak charges, $e^Z = 1-8/3\ \hbox{sin}^2\theta_W$, $-1+4/3\ \hbox{sin}^2\theta_W$ for the $u$ and $d/s$ quarks, respectively.

\begin{figure} 
\resizebox{20.5pc}{!}{\includegraphics{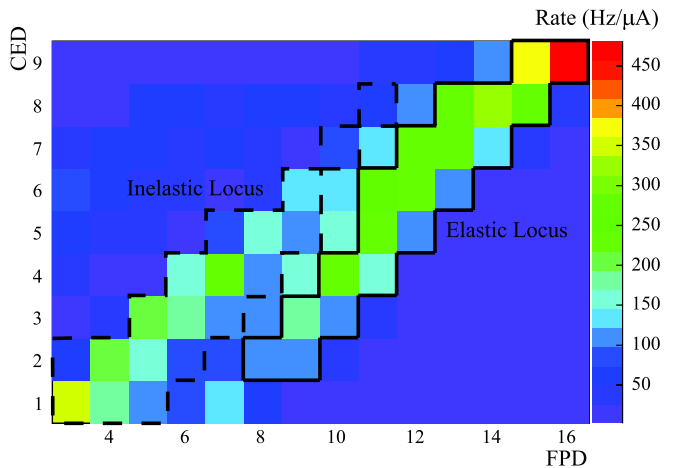}}
\caption{\label{fig:matrix} Example of counting rates -- LH$_2$, 0.684 GeV -- for various CED - FPD combinations (FPDs 1 and 2 not used).  Electrons from elastic (inelastic) scattering are in the upper right (lower left).}
\end{figure}

\begin{table}
\caption{\label{tab:backgrounds} Measured and raw elastic asymmetries (Eq. (\ref{eqn:background})).  $f$ is the background fraction for the dominant contribution (Al target cell) to the yield.  Misidentified $\pi^-$ contribute significantly only for the high $Q^2$ deuteron measurement with $f_{\pi} = 0.034\pm 0.010$.  $\Delta A_{corr}$ are the contributions to the overall point-to-point and global systematic uncertainties (Table \ref{tab:uncertainties}) due to these background corrections.}
\begin{ruledtabular}

\begin{tabular}{cccccc}
Target&$Q^2$&$A_{meas}$&$f$&$A_{el}$& $\Delta A_{corr}$\\
 &(GeV$^2$)& (ppm)& &(ppm)&(ppm)\\
\hline
H & 0.221 & -9.72 & $0.13 \pm 0.064$ & -9.22 & $\pm 0.11 \pm 0.40$\\
D & 0.221 & -13.50 & $0.099 \pm 0.050$ & -13.57 & $\pm 0.02 \pm 0.08$\\
H & 0.628 & -36.9 & $0.11 \pm 0.050$ & -37.0 & $\pm 0.61 \pm 0.86$\\
D & 0.628 & -37.4 & $0.061 \pm 0.031$ & -39.4 & $\pm 0.48 \pm 0.23$\\
\end{tabular}
\end{ruledtabular}
\end{table}

\begin{table*}
\caption{\label{tab:uncertainties} Corrections to the raw elastic asymmetries (Table \ref{tab:backgrounds}), and the resulting final physics asymmetries.  Rate and ``Other'' corrections are additive; beam polarization and electromagnetic radiative corrections are multiplicative.  ``Other'' corrections include those for helicity-correlated beam parameters, the small transverse component of beam polarization, and two-boson exchange.  The uncertainties for the corrections are point-to-point and global systematic; for the physics asymmetry the uncertainties are statistical, point-to-point and global systematic.}
\begin{ruledtabular}

\begin{tabular}{ccccccc}
Target & $Q^2$ & Rate & Other & Beam Polarization & EM Radiative & $A_{phys}$\\
 &(GeV$^2$)& (ppm) &(ppm) & & & (ppm) \\
\hline
H & 0.221 & $-0.31 \pm 0.08 \pm 0$ & $0.22 \pm 0.08 \pm 0.01$ & $\left (1/0.858 \right) \pm  0.02 \pm 0.01$ & $1.037 \pm 0.002 \pm 0$ & $-11.25 \pm 0.86 \pm 0.27 \pm 0.43$\\
D & 0.221 & $-0.58 \pm 0.21 \pm 0$ & $0.06 \pm 0.10 \pm 0.01$ & $\left (1/0.858 \right) \pm  0.02 \pm 0.01$ & $1.032 \pm 0.004 \pm 0$ & $-16.93 \pm 0.81 \pm 0.41 \pm 0.21$\\
H & 0.628 & $-1.28 \pm 0.18 \pm 0$ & $0.29 \pm 0.11 \pm 0.01$ & $\left (1/0.858 \right) \pm  0.01 \pm 0.01$ & $1.037 \pm 0.002 \pm 0$ & $-45.9 \pm 2.4 \pm 0.8 \pm 1.0$\\
D & 0.628 & $-7.0 \pm 1.8 \pm 0$ & $0.34 \pm 0.21 \pm 0.01$ & $\left (1/0.858 \right) \pm  0.01 \pm 0.01$ & $1.034 \pm 0.004 \pm 0$ & $-55.5 \pm 3.3 \pm 2.0 \pm 0.7$\\
\end{tabular}
\end{ruledtabular}
\end{table*}

Extracting $G_E^{p,Z}$ and $G_M^{p,Z}$ from the asymmetry, in turn, requires measurements at two different angles; in addition, a third measurement is necessary to determine the effective axial form factor, $G_A^e$.  The asymmetry in quasi-elastic scattering from the deuteron provides this independent combination of form factors. 

In addition to the strange quark vector currents, which are the main focus of this work, we present results for the isovector part of $G_A^e$, $G_A^{e,T=1}$, defined via
\begin{equation}
G_A^e = G_A^{e,T=1} + G_A^{e,T=0} = G_{A,cc}^{(0)} + R_{ana} + G_A^{e,T=0}.
\label{eqn:gae}
\end{equation}
To lowest order, it is the same as that measured in charged current neutrino scattering ($G_{A,cc}^{(0)}$)~\cite{beckmckeown,bernard,bodek}.  However, the radiative corrections, ($R_{ana}$), are expected to be significant ($\sim 30$\%) and distinct from those measured in neutrino scattering~\cite{zhu00a}.  They include the effect of the anapole moment, the effective parity-violating coupling of the photon to the nucleon~\cite{anapole}.
The isoscalar contribution to $G_A^e$, $G_A^{e,T=0}$, is a smaller ($< 10\%$)~\cite{zhu00a,HERMES}.  Our results (at $Q^2>0.1$ GeV$^2$) give the first indication of the $Q^2$ dependence of $G_A^{e,T=1}$.

We performed the G0 experiment~\cite{G0equip} in Hall C at Jefferson Lab.  We used polarized electron beams with currents up to $I = 60$ $\mu$A and energies of 359 and 684 MeV generated with a strained GaAs polarized source~\cite{poelker00a}.  The average beam polarization, measured with M\o ller and Mott~\cite{spindance} polarimeters, was $85.8 \pm 2.1 (1.4)\%$ at the lower (higher) incident energy.
Helicity-correlated current changes were corrected with active feedback to about 0.3~parts-per-million (ppm).  Corrections to the measured asymmetry for residual helicity-correlated beam current, position, angle and energy variations of $0.2 \pm 0.07$~ppm in the worst case are distributed roughly equally among these parameters and applied via linear regression.

A superconducting toroidal spectrometer, consisting of an eight-coil magnet, and eight detector sets, detected the electrons scattered at an angle of about 110$^\circ$ from 20 cm liquid hydrogen and deuterium targets~\cite{G0targ}.  Each detector set included two arrays of scintillators, one near the exit of the magnet (``CED''), and the second along its focal surface (``FPD'').  This combination of detectors allowed us to separate electrons from elastic and inelastic scattering (Fig.\ \ref{fig:matrix}).  An aerogel \v{C}erenkov detector with a pion threshold of 570 MeV, used in coincidence with the scintillators, allowed us to distinguish pions and electrons.  The largest pion to electron ratio (deuteron target at 684 MeV) was 5:1; the \v{C}erenkov detector had a rejection factor $\ge 85$ with an electron efficiency of about 85\%.  

Generically, the measured asymmetry has two components
\begin{equation}
A_{meas}=\left( 1-f \right) A_{el}+ f A_{b}
\label{eqn:background}
\end{equation}
where $A_{el}$ is the raw elastic asymmetry, $A_{b}$ the background asymmetry and $f$ the background fraction.  The backgrounds in the region of the elastic locus (see Fig.\ \ref{fig:matrix}) amount to 10-15\% of the signal.  In the elastic locus, the aluminum target windows dominate the backgrounds in the case of the proton and low-energy deuteron measurements; misidentified $\pi^-$ also contribute significantly for the high-energy deuteron measurement.  The aluminum fraction was measured using runs with gaseous hydrogen in the target (Table~\ref{tab:backgrounds}).  
The aluminum asymmetry was taken to be the same as that of the deuteron (both effectively quasi-elastic scattering only) with an additional uncertainty of 5\% for nuclear effects.  
The background corrections are small because the background asymmetries generally have values close to those of the elastic asymmetry. 

High speed scalers recorded the individual events for all CED-FPD pairs for both electrons and pions.  
All asymmetries were corrected for measured rate dependent effects (Table~\ref{tab:uncertainties}).  
For elastic scattering, dead-time corrections generally dominated those from accidentals and amounted to $\sim15\%$ of the yield based on the measured beam current dependence, and led to an uncertainty of about 0.5~ppm in the asymmetries.  
In the high-energy deuteron measurement, accidentals from pion signals in the scintillators in coincidence with random signals from the \v{C}erenkov dominated the correction.  In this case, the correction to the asymmetry was $-7.0 \pm 1.8$~ppm.  Electromagnetic radiative corrections~\cite{tsai} of $3 - 3.5\pm 0.3$\% and small two boson exchange effects (~1\%)~\cite{melnitchouk} were also applied to the asymmetries.
Table~\ref{tab:uncertainties} shows the corrections to the raw elastic asymmetry, $A_{el}$, as well as the final asymmetries $A_{phys}$ and their statistical and systematic uncertainties.

\begin{figure}
\resizebox{20pc}{!}{\includegraphics{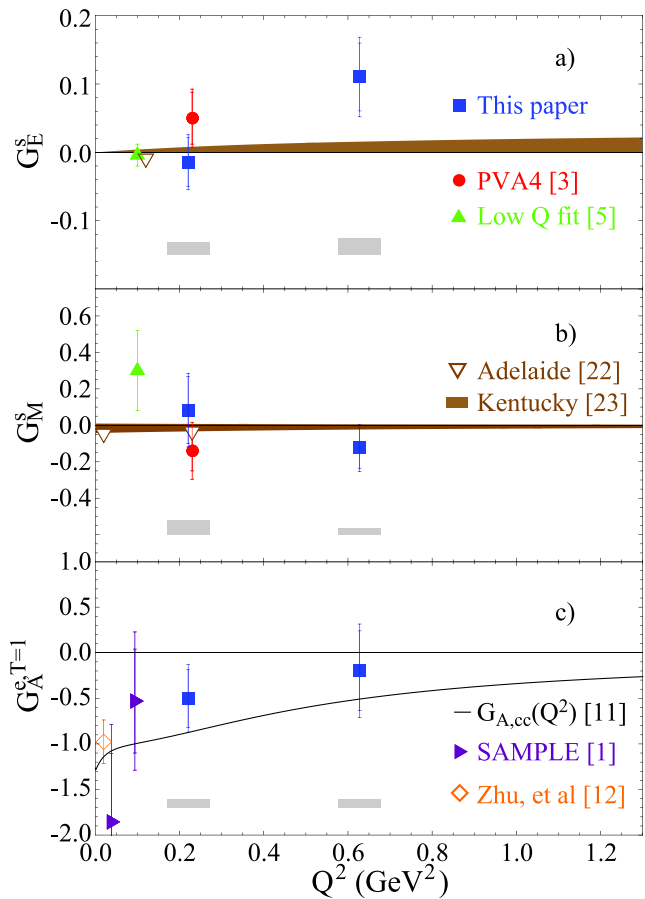}}
\caption{\label{fig:gesgmsgae}The form factors a) $G_E^s$, b) $G_M^s$, and c) $G_A^e$ determined by the G0 experiment forward- and backward-angle measurements.  Error bars show statistical and statistical plus point-to-point systematic uncertainties (added in quadrature); shaded bars below the corresponding points show global systematic uncertainties (for G0 points).  For $G_E^s$ and $G_M^s$, the extraction from Ref.~\cite{liu} as well as the results of the PVA4 (Mainz) experiment~\cite{mainz09} are shown.  Recent calculations from Adelaide~\cite{adelaide} and Kentucky~\cite{kentucky} groups are also shown; for the former the uncertainties are smaller than the symbols.  For $G_A^{e,T=1}$, results from the SAMPLE experiment~\cite{beisepittspayde} are shown together with the calculation of Zhu, et al.~\cite{zhu00a}.
}
\end{figure}

Fig.\ \ref{fig:gesgmsgae} shows the three new elastic form factors, $G_E^s$, $G_M^s$ and $G_A^{e,T=1}$, extracted from $A_{phys}$, at $Q^2 = 0.221$ and $0.628$ GeV$^2$~\cite{backweb}.  These results utilize a simple interpolation of our earlier forward angle measurements~\cite{forwardnote}. We have chosen the Kelly~\cite{kelly04} electromagnetic nucleon form factors, $G_{E,M}^{p,n}$, as the basis for these determinations to be consistent with our deuteron model~\cite{schiavilla}.  The isoscalar contributions to $G_A^e$ are taken from Refs.~\cite{zhu00a, HERMES}.  In addition to the experimental uncertainties already discussed, the point-to-point systematic uncertainties for the form factors
include contributions from 
the backward angle incident energies, four-momentum transfers, electromagnetic form factors and the deuteron model.  The largest contributions are from the momentum transfer and deuteron model, increasing this systematic uncertainty by about 10\% (relative to the total from $A_{phys}$).  The global uncertainties include contributions from the uncertainties in the forward angle incident energy, the electroweak radiative corrections~\cite{musolf94} and the isoscalar part of $G_A^e$.  In this case the largest contribution is from the electroweak radiative corrections and increases the global uncertainty by a few percent.

Fig.\ \ref{fig:gesgmsgae} also shows an extraction of $G_E^s$ and $G_M^s$ at $Q^2 = 0.1$ GeV$^2$ using a low $Q^2$ fit to previous data~\cite{liu}.  Lacking a backward angle deuteron measurement, the PVA4 points shown~\cite{mainz09}, in contrast to our results, {\it assume} a value for $G_A^{e,T=1}$ determined by the normalization of Ref.~\cite{zhu00a} (shown in Fig.\ \ref{fig:gesgmsgae}c), and a dipole form factor with a mass parameter of $1.032$ GeV.  The determinations of $G_A^{e,T=1}$ in the SAMPLE experiments~\cite{beisepittspayde} assume $G_M^s = 0.23\pm 0.36\pm 0.40$.  The contributions from both $G_E^s$ and $G_M^s$ in the SAMPLE measurements are small relative to the uncertainties.

The results indicate that strange quarks make small ($\lesssim 10$\%) contributions to the ground state charge and magnetic form factors of the nucleon.  Although the total $s$ quark momentum measured in deep-inelastic scattering is approximately one half that of $u$ and $d$ sea quarks~\cite{NuTeVsx}, our results suggest no significant spatial separation of $s$ and $\bar s$, consistent with the small differences in their measured momentum distributions~\cite{CTEQ}. 
The positive value of $G_E^s$ at $Q^2 = 0.628$ GeV$^2$ reflects the systematically positive values of the quantity $G_E^s + \eta G_M^s$ observed in the forward angle G0 measurements~\cite{forward}.  The values of $G_A^e$ reported here give the first experimental indication of the $Q^2$ dependence of the nucleon anapole moment effects~\cite{riska,maekawa}.

In summary, we have measured backward angle parity-violating asymmetries in elastic electron-proton and quasi-elastic electron-deuteron scattering at $Q^2 = 0.221$ and $0.628$ GeV$^2$.  These asymmetries determine the neutral weak interaction analogs of the ordinary charge and magnetic form factors of the nucleon, together with the effective axial form factor.  From the asymmetries we have determined $G_E^s$, $G_M^s$ and $G_A^{e,T=1}$, which indicate that the strange quark contributions to the nucleon form factors are $\lesssim 10$\%, and provide the first information on the $Q^2$ dependence of $G_A^{e,T=1}$.  Future forward angle experiments at $Q^2=0.63$ GeV$^2$ at Jefferson Lab and Mainz will further improve the precision of these determinations.

\begin{acknowledgments}
We gratefully acknowledge the strong technical contributions to this experiment from many groups: Caltech, Illinois, LPSC-Grenoble, IPN-Orsay, TRIUMF and particularly the Accelerator and Hall C groups at Jefferson Lab.  CNRS (France), DOE (U.S.), NSERC (Canada) and NSF (U.S.) supported this work in part.
\end{acknowledgments}


\begin{thebibliography}{27}
\bibitem[{()}]{forwardnote}
\bibinfo{note}{From Ref.~\cite{forward}: A linear fit was made to $A_{phys}-A_{NVS}$ values (see text) from $Q^2 = 0.177$ to $0.997$ GeV$^2$; the uncertainty of the interpolated values taken to be 70\% of the statistical uncertainty at the nearest measured point.}
\end{thebibliography}
\end{document}